# Simulation dynamique du voilier

Kostia RONCIN, Jean-Michel KOBUS

Ecole Centrale de Nantes - Laboratoire de Mécanique des Fluides.

1 rue de la Noë - B.P. 92101 - 44321 Nantes cedex 3

**Abstract:** A sailing simulator has been developed from a new point of view. It will rather be for the sailor to use it than for the architect. One boat which characteristics were already known has been chosen. The whole coupled mechanic equations system has been solved. Heave and pitch equations, usually neglected, have been introduced to evaluate each trim of the crew and provide an optimal ride analysis tool. In this paper, efforts models and their determination's method are briefly presented. A tack simulation is shown as a brief overview on the simulator possibilities.

## INTRODUCTION

La prédiction des performances des voiliers notamment pour l'America's Cup a de longue date généré des études scientifiques. A l'origine ces études étaient naturellement orientées vers la conception ; les architectes cherchant à optimiser les caractéristiques des carènes et des voiles sous la contrainte des règles de jauge. Jusqu'à ces dernières années les programmes dénommés VPP

(Velocity Prediction Program) permettaient de déterminer les points de fonctionnement stationnaires avec 3 des 6 équations de la mécanique (traînée, portance, roulis). Les forces hydrodynamiques sur les carènes sont en général obtenues par des modèles élaborés à partir d'essais systématiques en bassin des carènes. [6] [16]. On relève cependant une évolution vers des études plus complexes et exhaustives qui vont de pair avec des mesures en bassin et en soufflerie de plus en plus élaborées et l'utilisation de logiciel de calcul d'écoulement sur les carènes et les voiles. Les premiers simulateurs dynamiques [9] [10] ont introduit les équations de la manœuvrabilité avec le lacet comme quatrième degré de liberté. Les critères d'optimisation s'affinent, en 1987 pour la campagne du bateau Stars and Stripes un programme de simulation sur un parcours de régate complet a été réalisé, le critère retenu étant le temps de parcours [13]. En 1997 Caponnetto [2] introduit la présence d'un adversaire en prenant en compte l'interaction aérodynamique au prés entre voiliers.

Notre démarche s'inscrit dans cette évolution vers la complexité mais elle a la particularité de s'attacher à des voiliers de compétition monotypes dont les caractéristiques sont connues. L'objectif principal est d'approfondir la connaissance du comportement dynamique d'un voilier donné pour permettre d'en optimiser la conduite. Cette étude s'oriente d'avantage vers le sportif que vers le concepteur. Le simulateur devra permettre l'étude de l'influence des paramètres sur lesquels peut effectivement jouer l'équipage en compétition. Ce parti pris nous impose de tenir compte de l'ensemble des six degrés de liberté dont en particulier le pilonnement et le tangage jusque là négligés. Les essais que nous avons réalisés montrent en effet que l'influence de l'assiette sur la traînée peut être de plus de 20%. Pour un monotype de 8m comme le First Class 8, à vitesse élevée l'effort hydrodynamique vertical enfonce le bateau de façon considérable, l'attitude s'en trouve également modifiée. Le phénomène, largement couplé et non linéaire affecte la résistance à l'avancement. Ces résultats confortent l'idée que le système complet doit être résolu pour une représentation fiable du comportement dynamique complet du voilier.

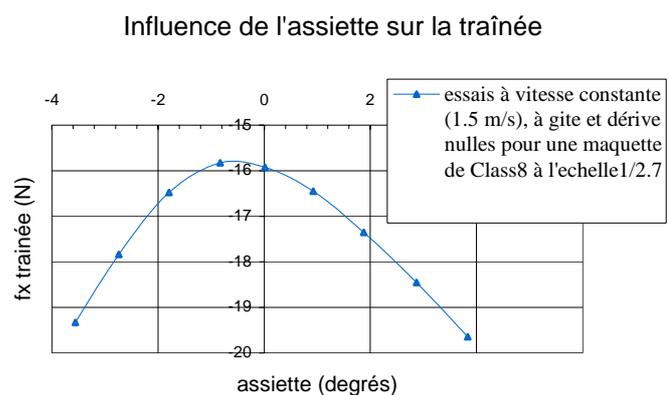

Les études précédentes sur le sujet ont montré l'importance prépondérante des efforts hydrodynamiques sur la performance des voiliers, c'est donc par l'évaluation de ces efforts que nous avons commencé notre étude. Avec les moyens expérimentaux dont nous disposons nous avons déterminé l'influence de chaque paramètre d'attitude sur les efforts hydrodynamiques. Après une brève présentation des équations de la dynamique et des hypothèses simplificatrices adoptées nous proposons une description des modèles d'efforts adoptés.

**NOTATIONS, REPERES ET ANGLES**

$\vec{V}_B$ : vitesse du centre de gravité du voilier dans le référentiel absolu ($R_0$).

VMG (Velocity made good) : projection de $\vec{V}_B$ sur l'axe du vent réel.

$C_L$ : coefficient de portance.  $V_{WA}$ : vitesse du vent apparent.
$C_D$ : coefficient de traînée.  $V_{WT}$ : vitesse du vent réel.
$C_T$ : coefficient d'effort total.  $\beta_{WA}$ : angle du vent apparent.
$\rho_A$ : masse volumique de l'air.  $\gamma$ : ratio traînée sur portance pour le gréement (voiles, mat, etc.).
  $\delta$ : angle de barre

## Référentiels

$R_0$ est un repère galiléen lié à la terre. Son origine est la surface libre au repos, l'axe $X_0$ est dans la direction du nord géographique, $Z_0$ est dirigé vers le haut. $R_B$ est le repère lié au bateau, déduit par translation du repère de définition de la carène, son origine est au centre de gravité nominal. $X_B$ est dirigé vers l'avant, $Y_B$ vers bâbord, $Z_B$ vers le haut. Pour les problèmes de manœuvrabilité l'habitude veut que l'on prenne $Z_B$ dirigé vers le bas et $Y_B$ vers tribord. Pour ne pas multiplier les référentiels nous préférons reprendre le repère $R_B$.

## Attitude et position

Le voilier est assimilé à un solide dont la position est connue à tout instant par le vecteur position et les trois angles de cardan, la gîte $\phi$, l'assiette $\theta$, le lacet $\psi$.

Le vecteur position : 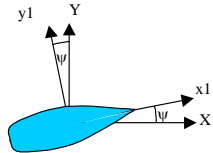 $\vec{P} = \overrightarrow{O_0 O_b} = \begin{pmatrix} x \\ y \\ z \end{pmatrix}$

L'attitude est : 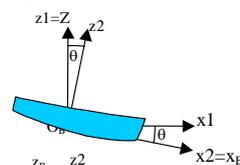 $\begin{pmatrix} \phi \\ \theta \\ \psi \end{pmatrix}$ et le vecteur rotation instantané dans ($R_B$) s'exprime par :

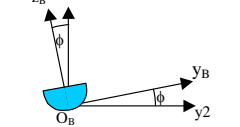 $\begin{pmatrix} p \\ q \\ r \end{pmatrix} = [\Sigma] \bullet \begin{pmatrix} \dot{\phi} \\ \dot{\theta} \\ \dot{\psi} \end{pmatrix}$ ; avec $[\Sigma] = \begin{bmatrix} 1 & 0 & -\sin\theta \\ 0 & \cos\phi & \cos\theta * \sin\phi \\ 0 & -\sin\phi & \cos\theta * \cos\phi \end{bmatrix}$

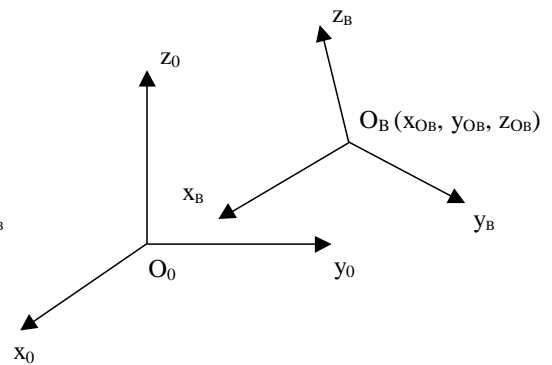

## METHODE POUR LA REALISATION DU SIMULATEUR

Nous travaillons sous l'environnement Matlab-Simulink qui permet de dissocier dans le simulateur des entités indépendantes. Chaque entité est définie graphiquement par un objet, elle est reliée aux autres entités par des relations d'entrées-sorties. Ce mode de programmation très visuel est parfaitement adapté à l'objectif de modularité que nous nous sommes fixé, le but étant de disposer d'un outil qui puissent intégrer différents modèles, obtenus par des études numériques ou expérimentales au fur et à mesure de leurs évolutions.

## LES EQUATIONS DE LA DYNAMIQUE ET LA MODELISATION DES EFFORTS

Les équations sont écrites en $O_B$ que l'on place au centre de gravité du voilier. On assimile le voilier à un solide rigide. Le principe fondamental de la dynamique donne :

$$\left\{\begin{array}{c}\text{Torseur des}\\ \text{efforts}\\ \text{aérodynamiques}\end{array}\right\}_{O_B,R_B} + \left\{\begin{array}{c}\text{Torseur des}\\ \text{efforts}\\ \text{hydrodynamiques}\end{array}\right\}_{O_B,R_B} + \left\{\begin{array}{c}\text{Torseur des}\\ \text{efforts}\\ \text{de pesanteur}\end{array}\right\}_{O_B,R_B} = \left\{\begin{array}{c}\text{Torseur}\\ \text{dynamique}\end{array}\right\}_{O_B,R_B}$$

### EFFORTS AERODYNAMIQUES.

La modélisation des efforts aérodynamiques a longtemps été le parent pauvre des études sur les voiliers, probablement à cause des difficultés théoriques et expérimentales, mais aussi parce qu'une évaluation sommaire de ces efforts suffit à obtenir des performances brutes du voilier satisfaisantes. Récemment les études numériques se sont développées, d'abord en fluides parfaits [11] [3] [8] et plus récemment en fluide visqueux [3]. Si les méthodes "Fluide Parfait" évaluent correctement la portance au près serré, lorsque les voiles travaillent en finesse, il n'en va pas de même de la traînée qui est largement affectée par les décollements tourbillonnaires à la jonction du mat et de la grand voile. Si bien que les codes fluides parfaits ne peuvent être exploités pour évaluer la performance sans être associés à des résultats expérimentaux. On a donc cantonné ces méthodes soit à l'étude des déformations, soit à l'étude de l'interaction entre les voiles [11] [2]. A l'aide des nouveaux outils informatiques, on commence à envisager l'étude en fluide visqueux. Mais les temps de calcul sont importants [3]. Caponnetto et al. donnent un ordre de grandeur de 8 heures pour la convergence du calcul sur les 12 processeurs d'une station Silicon Graphic origin2000. Pour les premières versions du simulateur nous avons opté pour une solution simple et éprouvée. Nous avons adapté le modèle empirique de Myers datant de 1975 [12], et repris dans de nombreux VPP à travers le monde [17] [13].

La force latérale au près est définie par :

$$F_L = \frac{\rho_A \cdot V_{wA}^2}{2} \cdot A_S \cdot \frac{C_L}{\cos\gamma} \cdot \cos(\beta - \gamma) \cdot \cos\phi$$

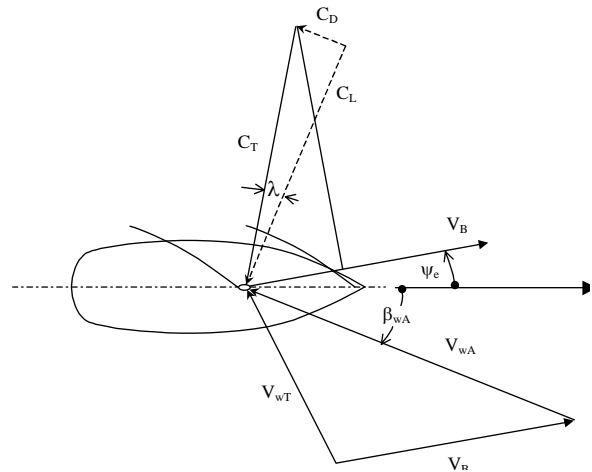

Myers donne une valeur indicative de 10° pour γ. Van Oossanen donne une évaluation systématique de $\gamma = 0.044 + \dfrac{0.089 \cdot C_L^2}{1-\phi}$

L'expression des coefficients de portance est déterminée par des formules de régression polynomiale au second ordre en fonction de l'angle d'incidence. Pour le génois, par exemple le modèle donne : $C_L = 1.04 + 0.344 \cdot \beta_{WA} + 0.041 \cdot \beta_{WA}^2$; $avec$ $\beta_{WA} \in \left[\dfrac{\pi}{8}; \dfrac{\pi}{2}\right]$ (génois)

Ce modèle ne fait pas intervenir de paramètres de réglages. Il est donc intéressant pour les premières versions d'un simulateur. En effet, l'intérêt d'introduire des réglages de voiles sophistiqués est discutable tant que les modèles utilisés dans les autres modules ne seront pas suffisamment fins pour que l'influence de ces réglages soit significative. Dans la littérature [1] [13] [14] [5 ], le réglage des voiles est introduit sous forme de deux coefficients, l'un pour réduire la voilure, il affecte donc la position du centre de poussée vélique et la surface de la voilure, tandis que le deuxième affecte les coefficients aérodynamiques pour tenir compte d'un réglage du creux des voiles. Ils pourront donc être introduits facilement dans le modèle par la suite.

**EFFORTS HYDRODYNAMIQUES**

Les efforts hydrodynamiques ont une évolution moins régulière, notamment la résistance de vagues, et doivent être évalués avec plus de soin, car la qualité de leur évaluation conditionne d'avantage la précision de la performance globale du voilier. Pour la plupart des modélisations nous souhaitons utiliser les outils expérimentaux et numériques développés au Laboratoire de Mécanique des Fluides de l'Ecole Centrale de Nantes. Le Laboratoire dispose en effet de moyens d'essais lourds en hydrodynamique, notamment un bassin des carènes de 72 m, un dynamomètre à 6 composantes et un orienteur spécifique pour l'étude des voiliers [15]. Des outils numériques de calcul d'écoulement autour des carènes, en code « Fluide Parfait » (REVA, Aquaplus [4]) et en code Navier-Stokes [1], sont également développés et perfectionnés par un travail de recherche permanent.

Actuellement, les efforts hydrodynamiques sur la carène en gîte et dérive à vitesse constante sont obtenus à partir de différentes méthodes d'interpolation entre des points de fonctionnements issus des campagnes d'essais en bassin des carènes. La méthode employée pour la répartition des points de fonctionnements est celle des plans d'expériences. Le problème générique est de modéliser les grandeurs mises en jeu (le torseur des efforts appliqués à la carène, aux voiles, aux œuvres mortes, etc.) à partir de la connaissance de quelques points de fonctionnements. Les outils d'interpolations courants (linéaire, splines, splines cubiques), permettent de résoudre ce problème à l'aide d'un maillage complet de l'espace vectoriel à n dimensions, où n est le nombre de facteurs qui influencent la grandeur étudiée. Pour nos modélisations le nombre de facteurs est tel, que réaliser un maillage complet du domaine étudié se révèle trop coûteux en regard des temps de calcul ou d'expérimentation, donc des coûts. La théorie des plans d'expériences, très utilisée aujourd'hui dans les milieux techniques et industriels, fournit outre une approche structurée, des outils d'optimisation du ratio informations recueillies sur le nombre d'expériences. Typiquement, notre objectif est de limiter le nombre de configurations d'essais à 128 par carène pour obtenir les efforts hydrodynamiques à vitesse constante dans toutes les attitudes possibles rencontrées en navigation. Cela représente approximativement 64 heures d'essais. Pour passer des résultats sur le modèle au réel, on suppose que la composante de pression des efforts est proportionnelle à $\lambda^3$, avec $\lambda$ le rapport d'échelle de longueur entre la maquette et le réel. On utilise la méthode préconisée par l'International Towing Tank Conference pour l'évaluation de la composante de friction et la méthode de Prohaska pour la détermination du coefficient de forme.

**EFFORTS DE PESANTEUR ET EFFORTS HYDROSTATIQUES**

Les efforts de pesanteur sont exprimés par leur résultante en $O_B$. La masse des équipiers et leurs positions est introduite indépendamment permettant le cas échéant de s'en servir comme un paramètre de réglage de l'attitude du bateau (gîte et assiette). Les efforts hydrostatiques sont obtenus par un modèle en fonction des paramètres d'attitude. Pour l'instant ce modèle est linéaire mais il n'y aura aucune difficulté à l'améliorer.

**EFFORTS INSTATIONNAIRES DE MANOEUVRABILITE, INERTIES AJOUTEES ET AMORTISSEMENTS**

Les efforts engendrés par les manœuvres du voilier dépendent de façon couplée de la position et des angles d'attitude avec leurs dérivées premières et secondes. Des formulations simplifiées ne prennent en compte que les coefficients des termes d'ordre le plus bas dans le développement de ces efforts. L'évaluation par le calcul des coefficients des termes de rotation n'est pas encore opérationnelle. L'évaluation expérimentale fait appel à des procédures et à du matériel très lourds dont nous ne disposons pas. Pour l'instant nous prenons en compte ces effets sur les surfaces portantes en transportant le torseur cinématique au centre d'effort de ces surfaces. Pour la carène nous utilisons pour l'instant des coefficients extrapolés de valeurs obtenues par Masuyama dans [1] pour un voilier de taille légèrement supérieure. Les inerties ajoutées sont également introduites de façon approximative mais une campagne de calcul avec le logiciel Aquaplus [4] permettra de les préciser. A terme, nous aurons la possibilité d'exploiter le code Icare [1] (fluide visqueux et mouvements quelconques) pour identifier tous les coefficients de manœuvrabilité sans réaliser d'expérimentation.

## RESULTATS

Nous présentons ici l'évolution de quelques paramètres significatifs du voilier pendant un virement de bord. La barre est commandée par un correcteur PID avec pour consigne une loi de cap. Cela ne correspond pas tout à fait à la manière d'opérer d'un barreur expérimenté mais ce résultat n'est qu'un exemple des possibilités offertes par le simulateur. On constate néanmoins que les réactions sont conformes aux observations des pratiquants. On observe la chute de la vitesse du bateau et de la vitesse de remontée au vent durant le virement. L'augmentation de la gîte en début de virement est également caractéristique.

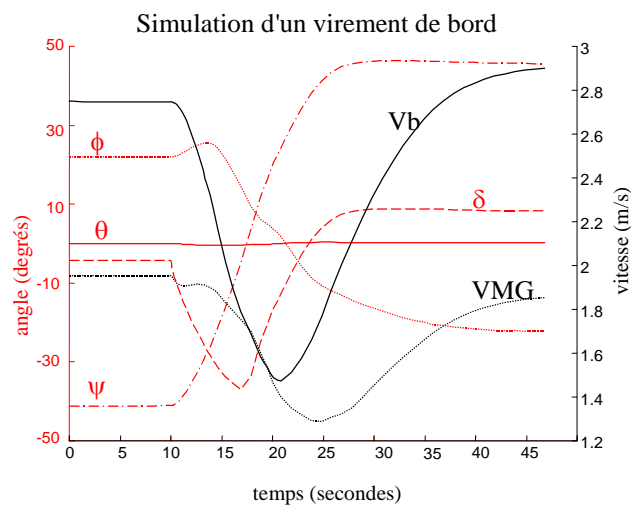

## CONCLUSION

Le simulateur que nous présentons est encore à l'état de prototype même si les premières exploitations semblent satisfaisantes du point de vue du fonctionnement général même s'il demande à être optimisé. Les premiers résultats montrent qu'il réagit de façon plausible. Sa structure modulaire permettra d'améliorer les modélisations grâce aux travaux en cours et aux résultats des recherches actuelles dans le domaine de l'hydrodynamique navale et de l'aérodynamique des voiles. Dans un bref avenir nous introduirons les efforts hydrostatiques non linéaires et les forces d'inertie ajoutées calculées. Mais la validation définitive viendra de l'expertise de barreurs confirmés auxquels nous demanderons de l'utiliser et surtout de mesures en mer à cap constant et en manœuvre. Pour les validations en mer, le déplacement des équipiers à bord sera pris en compte dans le simulateur pour rendre l'attitude du bateau en navigation et en manœuvre plus proche de la réalité. Nous participons actuellement à la mise en en place des collaborations qui nous permettrons de réaliser ces validations.